\documentclass[11pt]{article}
\title{ETH is Dead, Long Live SH}%
\author{%
  MIT Hardness Group%
\footnote{\protect\label{footnote:1}%
    Artificial first author to highlight that the other authors (in
    alphabetical order) worked as an equal group. Please include all
    authors (including this one) in your bibliography, and refer to the
    authors as “MIT Hardness Group” (without “et al.”).}
  \and
  Josh Brunner%
\footnote{\protect\label{footnote:MIT}%
    Computer Science and Artificial Intelligence Laboratory,
    Massachusetts Institute of Technology,
    32 Vassar St., Cambridge, MA 02139, USA.}
  \and
  Erik D. Demaine\footref{footnote:MIT}
  \and
  Fabian Frei\footref{footnote:MIT}%
\footnote{\protect\label{footnote:ETHZ}%
    Work done while not at ETH.
    }
  \and
  Timothy Gomez\footref{footnote:MIT}
  \and
  Markus Hecher%
\footnote{\protect\label{footnote:CNRS}%
    CNRS, UMR 8188, University of Artois, Centre de Recherche en Informatique de Lens (CRIL), France.}
  \and
  Jeffery Li\footref{footnote:MIT}
  \and
  Jayson Lynch\footref{footnote:MIT}
}%
\date{}

  \usepackage{amsmath}
  \usepackage{amssymb}
  \usepackage{amsthm}
  \usepackage[margin=1in]{geometry}
  \usepackage{microtype}
  \usepackage{tcolorbox}
  \usepackage{booktabs}
  \usepackage{tabularx}
  \usepackage[hyphens]{url}
  \usepackage[colorlinks=true,allcolors=blue]{hyperref}
  \usepackage{libertine}
  \usepackage[libertine]{newtxmath}
  \usepackage[scaled=0.96]{zi4}

  \makeatletter
  \long\def\linked@footnotetext#1#2#3{%
    \begingroup
      \def\@thefnmark{#1}%
      \def\Hy@footnote@currentHref{#2}%
      \let\@currentHref\Hy@footnote@currentHref
      \let\@currentlabelname\@empty
      \@footnotetext{#3}%
    \endgroup
  }
  \let\maketitle\HyOrg@maketitle
  \let\linked@makefnmark\@makefnmark
  \AddToHook{cmd/@maketitle/before}{\let\@makefnmark\linked@makefnmark}
  \renewcommand\thanks[1]{%
    \footnotemark
    \protected@xdef\@thanks{\@thanks
      \protect\linked@footnotetext{\@thefnmark}{\Hy@footnote@currentHref}{#1}}%
  }
  \let\Hy@tabularx@hook\@empty
  \long\def\TX@ftntext#1{%
    \edef\@tempa{\the\TX@ftn
      \noexpand\linked@footnotetext{\@thefnmark}{\Hy@footnote@currentHref}}%
    \global\TX@ftn\expandafter{\@tempa{#1}}%
  }
  \makeatother

\let\epsilon=\varepsilon
\newtheorem{theorem}{Theorem}[section]
\newtheorem{lemma}[theorem]{Lemma}
\newtheorem{corollary}[theorem]{Corollary}
\tcbset{%
  colback=violet!10,
  colframe=black,
  boxrule=0.5pt,
  arc=3pt,
  boxsep=0pt,
  left=8pt,
  right=8pt,
  top=6pt,
  bottom=6pt,
  before skip=\medskipamount,
  after skip=\medskipamount}%
\newtcolorbox{hypothesis}{}

\providecommand\defn[1]{\textbf{\textit{\boldmath #1}}}

\begin{document}
\maketitle
\begin{abstract}
  Among the recent flood of AI-based mathematical breakthroughs
  (the ``mathpocalypse''),
  perhaps the biggest bombshells in theoretical computer science
  were two disproofs of the Exponential Time Hypothesis (ETH),
  via slightly subexponential algorithms
  for the classic $k$-Satisfiability problems.
  After the initial shock, the community quickly began to wonder
  at the implications:
  how do the hundreds if not thousands of ETH-based lower bounds
  in the literature need to change?
  It turns out, not much --- if we are willing to change the exponents
  ever so slightly.

  While
  Impagliazzo and Paturi's 1999 Exponential Time Hypothesis (ETH) is false,
  Stearns and Hunt's 1990 Satisfiability Hypothesis (SH) remains plausible.
  We prove that SH is equivalent to the recently proposed Weak ETH
  which rules out $O(2^{n^{1-\epsilon}})$-time 3SAT algorithms
  for every constant $\epsilon>0$, where $n$ is the number of variables.
  We show that SH (or Weak ETH) preserves the typical process for lower bounds:
  a polynomial-time reduction from a 3SAT formula of size $s$
  (the number of variables plus clauses) to an instance of size
  $N\le s^{b+o(1)}$, for fixed blowup exponent $b>0$,
  rules out $O(2^{N^{1/b-\epsilon}})$-time algorithms for every constant
  $\epsilon>0$.
  By comparison, for an $N=O(s^b)$ blowup, ETH would have ruled out
  $2^{o(N^{1/b})}$-time algorithms.
  The framework also yields parameterized lower bounds ruling out a running time of
  $O(2^{k^{1/b-\epsilon}})L^{O(1)}$ when the reduction produces an instance
  of length $L$ with parameter $k\le s^{b+o(1)}$.
  We also prove equivalence of the analogous counting hypotheses,
  \#SH and Weak \#ETH, and that they rule out
  $O(2^{s^{1-\epsilon}})$-time algorithms
  for \#2SAT formulas of size $s$ for every constant $\epsilon>0$.
\end{abstract}

\section{Introduction}

Assuming $\mathrm{P}\ne\mathrm{NP}$, NP-complete problems cannot be solved in
$L^{O(1)}$ time, where $L$ is the length of the input.
But how quickly \emph{can} they be solved?
The obvious upper bound is $2^{L^{O(1)}}$ time,
by trying all possible polynomial-length certificates.
Stearns and Hunt~\cite{SH90} proposed measuring how fast an (exponential)
computational problem can be solved by its \defn{power index}:
the infimum over real powers $\alpha \geq 0$
for which the problem admits an $O(2^{L^{\alpha}})$-time algorithm
on inputs of length~$L$.%
\footnote{\protect\label{footnote:L}Although the original definition is in terms of \emph{bit} length,
  $L$ can count words of $L^{o(1)}$ bits each without changing the power index
  because $L^{\alpha} L^{o(1)} = O(L^{\alpha+\delta})$ for every fixed $\delta>0$.
  Placing the big $O$ in the exponent also leaves the power index unchanged:
  for any fixed $c,\delta>0$, $cL^{\alpha}\le L^{\alpha+\delta}$ for sufficiently
  large $L$.}

Ever since SAT, CNF SAT, and 3SAT were introduced as some of the first
NP-complete problems \cite{Coo71,Kar72,Lev73},
meaning that every NP problem reduces to them in polynomial time,
these problems have been a natural focus for faster algorithms.
The obvious algorithm for SAT checks all $2^n$ assignments to the $n$ variables
in $O(2^n L)$ time, which gives it a power index of at most~$1$.
In 1990, Stearns and Hunt \cite{SH90} conjectured that
CNF SAT's power index is exactly $1$:

\begin{hypothesis}
  \defn{Satisfiability Hypothesis (SH)} \cite{SH90}.
  For every constant $\epsilon>0$, CNF SAT on inputs of length $L$
  cannot be solved in $O(2^{L^{1-\epsilon}})$ time.\footnotemark
\end{hypothesis}

\footnotetext{
  \label{fn:cnf-encoding}
  Here we measure CNF input length $L$ as the number $n$ of variables
  plus the number $m$ of clauses plus the total number of literals.
  We index variables by $\{0,\dots,n-1\}$, giving an $O(L)$-word encoding.
  We count unused variables in $n$, so satisfying-assignment counts
  use at most $L+1$ bits.
  We assume no repeated literals within clauses, so $L=O(n+km)$ for $k$CNF;
  and no duplicate clauses, so $m=O(n^k)$ for $k$CNF with fixed $k$.}%

In 1999, Impagliazzo and Paturi \cite{IP99,IP01} proposed the stronger
Exponential Time Hypothesis, which asserts that \emph{3}SAT requires \emph{exponential}
time in the number $n$ of \emph{variables}:

\begin{hypothesis}
  \defn{Exponential Time Hypothesis (ETH)} \cite{IP99,IP01}.
  There is a constant $\epsilon>0$ such that 3SAT on inputs with $n$ variables
  cannot be solved in $O(2^{\epsilon n})$ time.
\end{hypothesis}

Recent algorithms by OpenAI \cite{OpenAI26} and Anthropic \cite{Anthropic26}
refute ETH and its counting analog \#ETH:
for every fixed $k$, they count the satisfying assignments of a $k$CNF formula
in
$$
  2^{O\bigl(n\log\log n/\sqrt{\log n}\bigr)}\text{ time}.
$$
The exponent $n\log\log n/\sqrt{\log n}$ is $o(n)$, which refutes ETH,
but is $n^{1-o(1)}$, which is consistent with~SH.

Thus we propose a return to the older SH hypothesis as a basis for lower bounds.
We show that many ETH-based lower bounds survive under SH,
with an arbitrarily small constant loss in the exponent.
This robustness stands in stark contrast to a recent \emph{WIRED} headline
declaring ``decades of computer science research `irrelevant'{}'' \cite{Ward26}.

To make SH easier to apply and more clearly related to ETH,
we consider its reformulation in terms of 3SAT and variable count,
recently introduced as ``Weak ETH''
\cite{Aar26,Lam26a,UCSD26,Sze26}.%
\footnote{\protect\label{footnote:weak-eth-terminology}%
  Manurangsi \cite{Man26} introduced the same hypothesis as the
  ``Almost-Exponential Time Hypothesis (Almost-ETH)'' in arXiv version~3,
  posted October~27, 2026, one day after the first ETH refutation
  was announced \cite{OpenAI26}.
  Earlier uses of ``Weak ETH'' differ.
  Lokshtanov \cite{Lok17} used the term for the weaker hypothesis that
  3SAT requires $2^{\Omega(n^c)}$ time for some $c>0$.
  Carmosino et al.\ \cite[Definition~32]{CHIKK21} use the term for a
  Minimum Circuit Size Problem (MCSP) hypothesis.
  Given this potential ambiguity, we recommend using SH beyond this paper.}

\begin{hypothesis}
  \defn{Weak ETH}.
  For every constant $\epsilon>0$, 3SAT on inputs with $n$ variables
  cannot be solved in $O(2^{n^{1-\epsilon}})$ time.
\end{hypothesis}

\subsection{Our Results}

\paragraph{Equivalent Hypotheses.}
First, we prove that SH is equivalent to Weak ETH
(Theorem~\ref{thm:sh-weak-eth} in Section~\ref{sec:equivalences}).
Thus we do not need a new assumption like Weak ETH
to handle the downfall of ETH;
the literature already had a suitable assumption, SH, from over 35 years ago.
Our proof first uses a linear CNF-to-3CNF reduction to relate input length
to $n+m$ (Lemma~\ref{lem:size}), then quantitative and
disjoint sparsification \cite{IPZ01,CIP06,DHMTW14} to connect size to variable count
(Lemma~\ref{lem:sparsification} and Corollary~\ref{cor:size}).
By taking care to make the reduction parsimonious,
the same proof establishes the corresponding equivalence for counting
versions of the hypotheses, \#SH and Weak \#ETH.

Thus we use SH (or the weaker \#SH) as our primary hypothesis,
and Weak ETH (or Weak \#ETH) as a convenient equivalent form
for analyzing reductions.

One consequence of our equivalence is that SH is equivalent to the
$k$SAT version of Weak ETH for every fixed $k\ge3$
(Corollary~\ref{cor:fixed-width}), and therefore these fixed-width versions of
Weak ETH are equivalent to each other,
as previously shown by the ETH Group at UC San Diego \cite{UCSD26}.

\paragraph{Reduction Framework.}
Our main result is a unified framework for deriving SH/Weak ETH-style
lower bounds for other computational problems via \emph{reductions},
analogous to how we used to use ETH for ETH-style lower bounds
(Section~\ref{sec:transfer}).
Specifically, a polynomial-time reduction from 3SAT with $n$ variables
and $m$ clauses to a target problem with output size at most $(n+m)^{b+o(1)}$,
for a blowup constant $b>0$,
implies that the problem has no $2^{O(L^{1/b-\epsilon})}$-time algorithms,
for every constant $\epsilon>0$, assuming SH (Theorem~\ref{thm:transfer}).
The $o(1)$ term permits polylogarithmic factors in the blowup.
The theorem also applies to subexponential reductions taking time
$2^{O((n+m)^{\alpha})}$ for any fixed $0<\alpha<1$, such as the reduction
for Independent Set parameterized by rank-width \cite{BKN23}.
Our framework combines the equivalences above with the power-index
transfer rule of Stearns and Hunt \cite[Theorem~2.4]{SH90}.

Table~\ref{tab:bounds} (top group) illustrates some lower bounds
resulting from applying our framework to known reductions.
Lampis \cite{Lam26a} made a similar observation that many ETH-based lower bounds
survive under Weak ETH with an arbitrarily small constant loss in the exponent;
our framework formalizes this observation.


\begin{table}[!tp]
  \centering
  \small
  \renewcommand{\arraystretch}{1.15}
  \begin{minipage} {\linewidth}
        \begin{tabularx}{\linewidth}{>{\raggedright\arraybackslash}Xccc}
    \toprule
    \textbf{Problem} & \textbf{Blowup} & \textbf{Under ETH} & \textbf{Under SH} \\
    \midrule
    3SAT \cite{IP01} & $s$ & $2^{o(n)}$ & $2^{O(n^{1-\epsilon})}$ \\
    Positive 1-in-3SAT \mbox{\cite[Thm.~2.1]{Sch78}}%
\footnote{\protect\label{footnote:positive-sat}%
      Positive means all literals are unnegated. 1-in-3 and NAE clauses require
      exactly one true variable and not all variables equal, respectively.
      These bounds hold even when every clause contains exactly three distinct
      variables. For the two nonplanar variants, apply Corollary~\ref{cor:schaefer}
      and separate any repeated variables using constant-size equality gadgets.}%
 & $O(s)$ & $2^{o(n)}$ & $2^{O(n^{1-\epsilon})}$ \\
    Positive NAE 3SAT \mbox{\cite[Thm.~2.1]{Sch78}} & $O(s)$ & $2^{o(n)}$ & $2^{O(n^{1-\epsilon})}$ \\
    Vertex Cover, Independent Set, 3-Coloring \mbox{\cite[Thms.~3.2--3.3]{LMS11}} & $O(s)$ & $2^{o(n)}L^{O(1)}$ & $2^{O(n^{1-\epsilon})}L^{O(1)}$ \\
    Treewidth \mbox{\cite[Thm.~1.2]{Bon26}}%
\footnote{\protect\label{footnote:bounded-occurrence}%
      Bounded-occurrence 3SAT is obtained with linear blowup by splitting variable occurrences and enforcing equality \cite[Proposition~3.1]{BBKMSZ20}.}%
 & $O(s)$ & $2^{o(n)}L^{O(1)}$ & $2^{O(n^{1-\epsilon})}L^{O(1)}$ \\
    Graph Homomorphism \mbox{\cite[Thm.~4.1]{CFGK17}} & $O(s/\log s)$ & $2^{o(n\log n)}L^{O(1)}$ & $2^{O(n^{1-\epsilon})}L^{O(1)}$ \\
    Planar 3SAT \mbox{\cite[Thm.~1]{Lic82}} & $O(s^2)$ & $2^{o(\sqrt{n})}$ & $2^{O(n^{1/2-\epsilon})}$ \\
    Planar Positive 1-in-3SAT \mbox{\cite[Prop.~3.3]{MR08}}%
\footnote{\protect\label{footnote:planar-positive-sat}%
      Compose Lichtenstein's $O(s^2)$-blowup planarization \cite{Lic82} with
      Mulzer and Rote's linear-blowup reduction from Planar 3SAT
      to Planar Positive 1-in-3SAT.}%
 & $O(s^2)$ & $2^{o(\sqrt{n})}$ & $2^{O(n^{1/2-\epsilon})}$ \\
    Planar Hamiltonian Cycle \mbox{\cite[Thm.~3.5]{LMS11}} & $O(s^2)$ & $2^{o(\sqrt{n})}L^{O(1)}$ & $2^{O(n^{1/2-\epsilon})}L^{O(1)}$ \\
    Subgraph Isomorphism on forests \mbox{\cite[Thm.~4]{BZ17}} & $O(s\log s)$ & $2^{o(n/\log n)}L^{O(1)}$ & $2^{O(n^{1-\epsilon})}L^{O(1)}$ \\
    Terrain Guarding, Orthogonal Terrain Guarding \mbox{\cite[Thm.~5]{BG19}} & $O(s^3)$ & $2^{o(n^{1/3})}L^{O(1)}$ & $2^{O(n^{1/3-\epsilon})}L^{O(1)}$ \\
    Dominating Set, Hamiltonian Cycle on induced $d$-dimensional grid graphs \mbox{\cite[Thms.~4.1, 4.7]{BBKMSZ20}}\footref{footnote:bounded-occurrence} & $O(s^{d/(d-1)})$ & $2^{o(n^{1-1/d})}L^{O(1)}$ & $2^{O(n^{1-1/d-\epsilon})}L^{O(1)}$ \\
    Euclidean TSP in $\mathbb{R}^d$ \mbox{\cite[Cor.~4.8]{BBKMSZ20}}\footref{footnote:bounded-occurrence} & $O(s^{d/(d-1)})$ & $2^{o(n^{1-1/d})}L^{O(1)}$ & $2^{O(n^{1-1/d-\epsilon})}L^{O(1)}$ \\
    \midrule
    $k$-Vertex Cover \cite{CJ03}; \mbox{\cite[Thm.~4.1]{LMS11}} & $O(s)$ & $2^{o(k)}L^{O(1)}$ & $2^{O(k^{1-\epsilon})}L^{O(1)}$ \\
    Planar $k$-Vertex Cover \cite{CJ03}; \mbox{\cite[Thm.~4.3]{LMS11}} & $O(s^2)$ & $2^{o(\sqrt{k})}L^{O(1)}$ & $2^{O(k^{1/2-\epsilon})}L^{O(1)}$ \\
    $k\times k$ Clique \mbox{\cite[Thm.~2.4]{LMS18}} & $O(s/\log s)$ & $2^{o(k\log k)}L^{O(1)}$ & $2^{O(k^{1-\epsilon})}L^{O(1)}$ \\
    Chromatic Number (treewidth or vertex cover number $k$) \mbox{\cite[Prop.~2.1, Thm.~6.1]{LMS18}}%
\footnote{\protect\label{footnote:structural-parameters}%
      These rows use different reductions for the ETH and SH bounds.
      The cited paper obtains the stronger $k\log k$ ETH lower bounds using
      additional arguments, including enumeration over colorings for Chromatic Number
      and a reduction through Permutation Independent Set for Disjoint Paths
      \cite[Section~2.3]{LMS18}.
      For SH, $O(s)$ blowup suffices.
      For Chromatic Number, the linear reduction from 3SAT to 3-Coloring
      \cite[Proposition~2.1]{LMS18} gives $O(s)$ vertices,
      bounding both parameters.
      For Disjoint Paths, encode SAT variables as rows of
      $q\times q$ Hitting Set, with $q=O(s)$.
      Add a set containing the two truth-value choices in each row,
      and a set of satisfying choices for each clause.
      Then apply the cited polynomial-time reductions, which give treewidth $O(q)$.}%
 & $O(s)$ & $2^{o(k\log k)}L^{O(1)}$ & $2^{O(k^{1-\epsilon})}L^{O(1)}$ \\
    Disjoint Paths (treewidth $k$) \mbox{\cite[Thms.~5.1, 5.4]{LMS18}}\footref{footnote:structural-parameters} & $O(s)$ & $2^{o(k\log k)}L^{O(1)}$ & $2^{O(k^{1-\epsilon})}L^{O(1)}$ \\
    Independent Set (rank-width $k$) \mbox{\cite[Thm.~1.1]{BKN23}}%
\footnote{\protect\label{footnote:rank-width-construction}%
      The weighted construction \cite[Section~4]{BKN23} has linear rank-width
      $O(\sqrt s)$, takes $2^{O(\sqrt s)}$ time, and uses weights at most
      $2^{O(\sqrt s)}$. Replacing weighted vertices by false twins
      (their Lemma~4.12) removes weights within the same time bound,
      increasing linear rank-width by at most one.
      Apply Theorem~\ref{thm:transfer} with $\alpha=b=1/2$.}%
 & $O(\sqrt s)$ & $2^{o(k^2)}L^{O(1)}$ & $2^{O(k^{2-\epsilon})}L^{O(1)}$ \\
    \bottomrule
    \end{tabularx}

    \caption{\label{tab:bounds}
      Historical ETH bounds, and their SH counterparts
      that follow from our reduction framework applied to the cited reductions.
      Each running-time entry is ruled out under the hypothesis named in its column.
      The SH bounds hold for every constant $\epsilon>0$.
      The blowup column bounds the target $n$ or $k$ in terms of $s$,
      the source 3SAT variable count plus clause count.
      The listed blowups suffice for the SH bounds but do not always yield
      the strongest ETH bounds.\footref{footnote:structural-parameters}
      In the running-time bounds, $L$ is the input length; $n$ counts variables for SAT,
      vertices for graph and terrain problems (summed over all input graphs),
      and points for Euclidean TSP;
      $d \ge 2$ is fixed;
      and $k$ is the solution size, number of rows, or structural parameter
      indicated in the problem column.}%
  \end{minipage}
\end{table}


\paragraph{Parameterized Lower Bounds.}
Our main transfer theorem applies to parameterized complexity as well
(Section~\ref{sec:transfer}):
if a polynomial-time reduction from 3SAT bounds the target parameter by
$k\le(n+m)^{b+o(1)}$, for a constant $b>0$, then SH rules out
$2^{O(k^{1/b-\epsilon})}L^{O(1)}$-time algorithms for every constant
$\epsilon>0$, where $L$ is the input length (Theorem~\ref{thm:transfer}).
Table~\ref{tab:bounds} (bottom group) illustrates some lower bounds
resulting from applying our framework to known parameterized reductions.

These lower bounds cannot distinguish logarithmic factors in the exponent,
however \cite{Lam26a,Lam26b}.
Szeider \cite{Sze26} proposes a separate \mbox{$k\times k$} Clique Hypothesis
to retain stronger parameterized lower bounds.

\paragraph{Boolean Constraint Satisfaction.}
Combining our reduction framework with our fixed-width equivalences,
we show that SH is equivalent to the variable-based lower bound for every
satisfiability problem on the NP-complete side of Schaefer's dichotomy
(Corollary~\ref{cor:schaefer} in Section~\ref{sec:schaefer}).

\paragraph{Approximation Lower Bounds.}
Our framework also applies to gap reductions (Section~\ref{sec:approximation}).
In particular, we rederive under SH Manurangsi's \cite[Section~2.2]{Man26}
constant-gap hardness for Max-3SAT (Corollary~\ref{cor:gap}) and their
approximation lower bound for Densest $k$-Subgraph \cite[Theorem~1]{Man26}.
The former uses Dinur's nearly linear PCP theorem \cite{Din07},
while the latter uses subexponential reductions.

\paragraph{Counting Lower Bounds.}
In Section~\ref{sec:counting}, we use linear-size reductions of Bannach, Demaine, Gomez,
and Hecher \cite{BDGH25,BDGH26}
(or a quantitative adaptation of Curticapean's earlier block interpolation \cite{Cur15})
to rule out
$2^{O((n+m)^{1-\epsilon})}$-time algorithms for \#2SAT under the weaker \#SH,
for every constant $\epsilon>0$ (Corollary~\ref{cor:counting-2sat}).
This application uses two counting queries and subtraction (or a single counting query and arithmetic postprocessing).

\paragraph{Randomization.}
Algorithms are deterministic unless specified otherwise.
In Section~\ref{sec:randomization}, we extend our equivalences and lower bounds
to bounded-error randomized algorithms under the corresponding randomized
hypotheses (Theorem~\ref{thm:randomized}).

\paragraph{Line of Succession.}
Finally, Section~\ref{sec:succession} discusses weaker hypotheses if SH should fail,
including positive power index and restrictions on quasipolynomial-time
algorithms.

\subsection{Alternative Hypotheses}


Cygan et al.\ \cite[Section~1]{CDL+16} propose the hypothesis that,
for every constant $\epsilon>0$, \emph{CNF} SAT on $n$ variables and $m$ clauses
cannot be solved in $O(2^{(1-\epsilon)n})m^{O(1)}$ time.
This hypothesis remains open after the recent fixed-width ETH refutations.
The polynomial factor $m^{O(1)}$ is essential:
$m$ can be exponential in $n$ even without duplicate clauses,
so omitting it would make the hypothesis follow
merely from the cost of reading the input.%
\footnote{\protect\label{footnote:rep}%
  For the same reason, we need to exclude repeated literals within clauses,
  as in footnote~\ref{fn:cnf-encoding}.}
However, as they note, the lack of a suitable sparsification lemma for
unbounded-width CNF prevents the usual transfer of such a hypothesis
to ETH-style lower bounds for other problems.

By contrast, Li~\cite[Hypothesis~2.14 and Section~6]{Li26} proposes
a family of CNF SAT hypotheses in terms of formula length $L$,
counted as the total number of literal occurrences
(which can be exponential in $n$ as mentioned above).
Parameterized by a constant $c>0$,
their hypothesis family rules out randomized $2^{L/(\log L)^c}$-time
algorithms for CNF SAT.
As Li notes, a linear-size conversion to 3CNF lets the recent algorithms solve
CNF SAT in $2^{O(L/\sqrt{\log L})}$ time.
Thus their hypothesis is false for $c<1/2$ and remains open for $c\ge1/2$.
Chen, Rothblum, Tell, and Yogev~\cite[Theorem~1.1]{CRTY23} earlier derived
derandomization assuming that, for some positive constant $c$, TQBF has no randomized
$2^{L/(\log L)^c}$-time algorithm, where $L$ is input bit length.
SH instead gives a single hypothesis without specifying a polylogarithmic rate.

\section{Hypothesis Equivalences \label{sec:equivalences}}

We prove that SH and Weak ETH are equivalent by relating input length,
formula size, and variable count.
At the same time, we prove equivalence of the corresponding counting
hypotheses, \#SH and Weak \#ETH, defined below following Dell et al.'s
\cite{DHMTW14} distinction between ETH and \#ETH.

\begin{hypothesis}
  \defn{Counting Satisfiability Hypothesis (\#SH)}.
  For every constant $\epsilon>0$, no algorithm counts
  the satisfying assignments of a CNF formula of length $L$ in
  $O(2^{L^{1-\epsilon}})$ time.
\end{hypothesis}

\begin{hypothesis}
  \defn{Weak \#ETH}.
  For every constant $\epsilon>0$, no algorithm counts
  the satisfying assignments of a 3CNF formula with $n$ variables
  in $O(2^{n^{1-\epsilon}})$ time.
\end{hypothesis}

SH implies \#SH because a positive count is equivalent to satisfiability.
Thus \#SH is a potentially weaker assumption.

\subsection{Input Length and Formula Size}

First we relate input length to the size $n+m$ of a fixed-width formula
using a parsimonious linear reduction. (Sparsification will then connect
this size to the number of variables.)
The decision case is implicit in Stearns and Hunt \cite[Section~3]{SH90};
we give a parsimonious proof to cover counting.

\begin{lemma} [Size Formulation of SH]\label{lem:size}
  For each fixed integer $k\ge 3$, SH is equivalent to the assertion that,
  for every constant $\epsilon>0$, $k$SAT has no
  $2^{O((n+m)^{1-\epsilon})}$-time algorithm.
  The analogous equivalence holds for \#SH and \#$k$SAT.
\end{lemma}

\begin{proof}
  We treat decision and counting together, decreasing any hypothetical
  power saving if necessary so that $0<\epsilon<1$.
  A $k$CNF formula has length $L=\Theta(n+m)$ for fixed $k$.
  Thus an algorithm violating SH or \#SH, restricted to $k$CNF, gives
  a $2^{O((n+m)^{1-\epsilon})}$-time algorithm.
  Conversely, convert a CNF formula of length $L$ to 3CNF using the
  \defn{Tseitin encoding} \cite{Tse70}:

  \begin{enumerate}
  \item Represent each clause as a binary OR circuit.
  \item Introduce a fresh variable for each gate's output.
     For a gate with inputs $x,y$ and output $z$, enforce
     $z\leftrightarrow(x\lor y)$ using the clauses
     $(\neg x\lor z)$, $(\neg y\lor z)$, and $(x\lor y\lor\neg z)$.
  \item Add a unit clause requiring each circuit's output to be true.
  \item Remove any repeated literals and duplicate clauses, preserving
     the satisfying assignments.
  \end{enumerate}

  This encoding uses $O(L)$ variables and clauses, each clause having at most
  three literals. Every assignment to the original variables uniquely
  determines the auxiliary gate values; the resulting extension satisfies
  the encoding if and only if the original assignment satisfies the
  original formula. Thus the reduction preserves both satisfiability
  and the number of satisfying assignments.
  The output is also a $k$CNF formula. A hypothetical algorithm for
  $k$SAT or \#$k$SAT with the claimed running time therefore solves
  the corresponding CNF problem in
  $2^{O(L^{1-\epsilon})}L^{O(1)}=O(2^{L^{1-\epsilon/2}})$ time,
  contradicting SH or \#SH.
\end{proof}

Thus restricting SH or \#SH to any fixed clause width $k\ge3$ gives an
equivalent hypothesis; see Corollary~\ref{cor:fixed-width} below.

\subsection{Sparsification and Variable Count}

Now we show that, for fixed clause width, lower bounds in terms of $n+m$
are equivalent to those in terms of $n$ (Corollary~\ref{cor:size}).
Manurangsi \cite[Lemma~4 and Appendix~B]{Man26} proved this equivalence
for 3SAT using quantitative sparsification with a subconstant parameter.
The ETH Group at UC San Diego \cite[Sections~2--5]{UCSD26} extended the
argument to every fixed clause width $k\ge3$.
We include these arguments for self-containedness and extend them to counting
using disjoint sparsification.

The key tool is quantitative sparsification.
For a CNF formula $F$, let $\#F$ denote the number of satisfying assignments
over its explicitly specified original variable set.

\begin{lemma} [Quantitative Sparsification {\cite{IPZ01,CIP06,DHMTW14}}]\label{lem:sparsification}
  For any fixed $k\ge 3$, there is a constant $C_k\ge 1$ such that,
  for every fixed $0<\delta<1$, a $k$CNF formula $F$ on $n$ variables
  and $m$ clauses can be expressed as
  $$ F = \bigvee_{i=1}^t F_i, \qquad t= 2^{O(n^{1-\delta})}, $$
  where every $F_i$ is a $k$CNF formula on the same original variables with
  $O(n^{1+C_k\delta})$ clauses.
  The satisfying assignments of the $F_i$ are pairwise disjoint, so
  $\#F=\sum_i\#F_i$.
  The formulas can be generated in
  $2^{O(n^{1-\delta})}(n+m)^{O(1)}$ time.
\end{lemma}

In the usual ETH argument, choosing a fixed
sparsification parameter $\eta$ below the ETH constant $\epsilon$ lets us combine
$2^{\eta n}$ branches with a hypothetical $2^{o(n+m)}$-time algorithm
on each sparse formula to beat the forbidden $2^{\epsilon n}$
threshold.
For Weak ETH, any fixed $\eta>0$ incurs too much overhead to obtain a
constant power saving in the exponent, so we instead let $\eta$ decrease
with $n$ and track how the output clause density grows.

\begin{proof}
  We follow the parameter choice in \cite[Appendix~B, Lemma~11]{Man26}.
  Sparsification with parameter $\eta$ produces at most $2^{\eta n}$
  formulas, each with $n(k/\eta)^{O(k)}$ clauses \cite[Section~6]{CIP06}.
  Set $\eta=n^{-\delta}$ and enlarge $C_k$ to absorb the constant
  in $O(k)$.
  To justify the claimed running time when $\eta$ depends on $n$,
  we inspect the sparsification algorithm.
  The algorithm branches on a subclause shared by sufficiently many
  clauses of the same size (which form an admissible ``flower'').
  For fixed $k$, these occurrence thresholds are polynomial in $1/\eta$,
  and suitable subclauses can be found by enumerating the
  $O(n^k)$ possible subclauses and counting their occurrences.
  The work along a branch is therefore polynomial in $n+m+1/\eta$,
  giving the claimed total time even when $\eta$ depends on $n$.
  The disjoint version of Dell et al.\
  \cite[Appendix~A, Lemma~A.1]{DHMTW14} introduces no additional
  branching and adds only $O(n)$ unit clauses per output formula,
  preserving these bounds.
\end{proof}

For a fixed integer $k\ge3$, define \defn{Weak $k$-ETH} by replacing 3SAT
with $k$SAT in Weak ETH: for every constant $\epsilon>0$, $k$SAT on inputs
with $n$ variables cannot be solved in
$O(2^{n^{1-\epsilon}})$ time.

\begin{corollary} [Size Formulation of Weak ETH]\label{cor:size}
  For each fixed integer $k\ge 3$, Weak $k$-ETH is equivalent to the
  assertion that, for every constant $\epsilon>0$, $k$SAT has no
  $2^{O((n+m)^{1-\epsilon})}$-time algorithm.
  Likewise, Weak \#ETH is equivalent to the same assertion for \#3SAT.
\end{corollary}

The choice $\eta=n^{-\delta}$ trades fewer output formulas for more
clauses per formula. We choose $\delta>0$ small enough that both the
branching and the per-formula running-time exponents remain below $1$.

\begin{proof}
  Suppose such an algorithm exists, using $k=3$ in the counting case.
  By decreasing the power saving if necessary, assume $0<\epsilon<1$.
  Let $C=C_k$ from Lemma~\ref{lem:sparsification} and choose a constant
  $0<\delta<1$ sufficiently small that $(1+C\delta)(1-\epsilon)<1$.
  Sparsify and run the algorithm on every resulting formula.
  For decision, take the disjunction of the answers; for counting,
  sum the counts, using disjointness.
  Each count has a binary representation of at most $n+1$ bits, so
  summation takes only polynomial time per formula. In both cases, the total time is
  $$
    2^{O\bigl(n^{1-\delta}+n^{(1+C\delta)(1-\epsilon)}\bigr)}
    (n+m)^{O(1)},
  $$
  which is $O(2^{n^{(1+\rho)/2}})$, absorbing the polynomial factor
  using $m=O(n^k)$, for
  $\rho=\max\{1-\delta,(1+C\delta)(1-\epsilon)\}<1$,
  contradicting Weak $k$-ETH or Weak \#ETH with power saving $(1-\rho)/2$.
  Conversely, an algorithm violating either variable-based hypothesis
  gives such a size-based algorithm because $n\le n+m$.
\end{proof}

\subsection{Equivalent Hypotheses}

Combining the size and variable formulations gives the desired equivalences.

\begin{theorem} [Decision and Counting Equivalences]\label{thm:sh-weak-eth}
  SH is equivalent to Weak ETH,
  and \#SH is equivalent to Weak \#ETH.
\end{theorem}

\begin{proof}
  Combine Lemma~\ref{lem:size} and Corollary~\ref{cor:size} with $k=3$.
\end{proof}

The variable-based fixed-width equivalence is established in the
power-index formulation by the ETH Group at UC San Diego
\cite[Section~6, Equation~(12)]{UCSD26}.
The same two size and variable formulations also connect it to SH.

\begin{corollary} [Equivalence of Fixed Widths]\label{cor:fixed-width}
  For every fixed integer $k\ge 3$, SH, Weak ETH, and Weak $k$-ETH are equivalent.
\end{corollary}

\begin{proof}
  Apply Lemma~\ref{lem:size} and Corollary~\ref{cor:size} for this $k$, and
  Theorem~\ref{thm:sh-weak-eth} for $k=3$.
\end{proof}

The resulting power saving in terms of variables may depend on $k$.

\section{Reduction Framework \label{sec:transfer}}

Next we develop a framework for lower bounds under SH by tracking the size blowup and
running time of reductions.
We combine Stearns and Hunt's power-index transfer theorem
\cite[Theorem~2.4]{SH90} with our SAT equivalences, formulate a version
for parameterized problems, and derive approximation and counting lower bounds.
The size formulation of SH in Lemma~\ref{lem:size} lets us work directly
with the source size $n+m$.
For parameterized complexity, we measure the target instance $I$ in terms
of a parameter $N$, which need not bound $I$'s encoding length $L$;
the reduction's running time bounds $L$.

\begin{theorem} [Exponent Transfer]\label{thm:transfer}
  Suppose there is a many-one reduction from 3SAT to a problem $\Pi$
  that, on a formula with $n$ variables and $m$ clauses, takes
  $2^{O((n+m)^{\alpha})}$ time and produces an instance $I$ with parameter
  $N\le(n+m)^{b+o(1)}$, for constants $0<\alpha<1$ and $b>0$.
  Assuming SH, for every constant $\epsilon>0$,
  $\Pi$ has no algorithm running in
  $2^{O(N^{1/b-\epsilon})}L^{O(1)}$ time.
\end{theorem}

\begin{proof}
  Let $s=n+m$. By decreasing the power saving if necessary, assume
  $0<\epsilon<1/b$. The reduction produces an instance with
  $L\le 2^{O(s^{\alpha})}$, so any polynomial factor in $L$ is also
  $2^{O(s^{\alpha})}$.
  Substituting $N\le s^{b+o(1)}$ into the hypothetical algorithm's running
  time and adding the reduction time gives a 3SAT algorithm running in
  $$
    2^{O(s^{\alpha})}
      +2^{O\bigl(s^{b+o(1)}\bigr)^{1/b-\epsilon}}2^{O(s^{\alpha})}
    =2^{O\bigl(s^{\alpha}+s^{1-b\epsilon+o(1)}\bigr)}
    =2^{O\bigl(s^{\max\{\alpha,1-b\epsilon\}+o(1)}\bigr)}\text{ time}.
  $$
  Because $\max\{\alpha,1-b\epsilon\}<1$, the $o(1)$ term leaves a fixed
  power saving, contradicting Lemma~\ref{lem:size}.
\end{proof}

An NP-hardness proof often consists of a chain of
polynomial-time reductions starting ultimately from 3SAT.
If the reductions have blowup exponents $b_1,\ldots,b_r$ respectively, then
they compose into a reduction 
whose blowup exponent is the product $b_1\cdots b_r$.
Theorem~\ref{thm:transfer} then rules out
$2^{O(N^{1/(b_1\cdots b_r)-\epsilon})}L^{O(1)}$-time algorithms for the final
problem under SH, for every constant $\epsilon>0$.

Table~\ref{tab:bounds} gives example applications of the framework.

The $\epsilon$ loss absorbs logarithmic factors in the exponent, so the
framework conflates historical ETH-based lower bounds such as $2^{o(N)}$
and $2^{o(N\log N)}$: both become $2^{O(N^{1-\epsilon})}$ bounds under SH.
An illustration of this loss is the \defn{$k\times k$ Clique} problem:
given a graph on $\{1,\ldots,k\}\times\{1,\ldots,k\}$,
decide whether there is a clique with one vertex from each row.
The polynomial-time reduction of Lokshtanov, Marx, and Saurabh
\cite[Proposition~2.1, Theorem~2.4]{LMS18} has
$k\log k=O(n+m)$, so Theorem~\ref{thm:transfer} with $N=k$ and $b=1$ rules out
$2^{O(k^{1-\epsilon})}L^{O(1)}$-time algorithms under SH for every constant
$\epsilon>0$, as in Table~\ref{tab:bounds}.
But a hypothetical $2^{O(k)}L^{O(1)}$-time algorithm, ruled out under ETH,
would give only a $2^{O((n+m)/\log(n+m))}$-time SAT algorithm,
which does not contradict SH.
Lampis \cite{Lam26a,Lam26b} pointed out this limitation of Weak ETH.
Szeider \cite{Sze26} proposes the separate \defn{$k\times k$ Clique Hypothesis}:
this problem has no $2^{o(k\log k)}L^{O(1)}$-time algorithm.
The recent breakthrough for $k$-clique \cite{VWZ26} does not affect this
hypothesis: applying the faster algorithm to $k \times k$ instances
improves only the constant in the exponent of a
$2^{\Theta(k\log k)}$ running time.

%

\section{Boolean Constraint Satisfaction \label{sec:schaefer}}

Combining our reduction framework with our fixed-width equivalences,
we show that every satisfiability problem on the NP-complete side of
Schaefer's dichotomy \cite{Sch78} yields an equivalent formulation of SH.

For a fixed finite set $\Gamma$ of Boolean relations,
the \defn{$\mathrm{SAT}(\Gamma)$} problem \cite{Sch78} asks whether
$n$ variables can be set to satisfy a given conjunction of $m$ clauses,
each of which applies a relation in $\Gamma$ to a tuple of the variables.
Variables within a clause may repeat.
We omit duplicate clauses, so $m = O(n^k)$, 
where $k$ is $\Gamma$'s maximum arity.
Schaefer \cite{Sch78} characterized the $\Gamma$ for which $\mathrm{SAT}(\Gamma)$
is polynomial-time solvable versus NP-complete; see \cite{Jea98} or
\cite[Sections~1.2.7--1.2.8]{DGH26} for modern treatments.

\begin{corollary} [Schaefer SAT Lower Bounds]\label{cor:schaefer}
  Let $\Gamma$ be a fixed finite set of Boolean relations on the NP-complete
  side of Schaefer's dichotomy \cite[Theorem~2.1]{Sch78}.
  SH is equivalent to the assertion that, for every constant $\epsilon>0$,
  $\mathrm{SAT}(\Gamma)$ on $n$ variables has no
  $O(2^{n^{1-\epsilon}})$-time algorithm.
\end{corollary}

\begin{proof}
  Schaefer's reduction from 3SAT to $\mathrm{SAT}(\Gamma)$
  \cite[Section~4]{Sch78}
  first replaces each 3SAT clause by a fixed-size gadget over~$\Gamma$,
  using fresh auxiliary variables and Boolean constants,
  and then removes the constants using fixed-size gadgets.
  Thus the reduction produces $O(n+m)$ variables
  and clauses from a 3SAT formula with $n$ variables and $m$ clauses.
  Theorem~\ref{thm:transfer} with $b=1$ gives the claimed lower bound assuming~SH.

  Conversely, we reduce $\mathrm{SAT}(\Gamma)$ to $k$SAT,
  where $k$ is $\Gamma$'s maximum arity.
  For each clause, forbid each violating assignment
  by adding a CNF clause containing each of its variables unnegated when assigned false
  and negated when assigned true.
  Because $\Gamma$ is fixed and finite, this reduction constructs a constant
  number of CNF clauses per original clause, on the same variables.
  Delete duplicate CNF clauses.
  If SH fails, Corollary~\ref{cor:fixed-width} gives an
  $O(2^{n^{1-\epsilon}})$-time algorithm for $k$SAT for some constant
  $\epsilon>0$. The reduction takes polynomial time in $n$, because
  there are no duplicate clauses. Absorbing this time by decreasing
  $\epsilon$ gives an algorithm violating the asserted lower bound.
\end{proof}

\section{Approximation Under SH \label{sec:approximation}}

A common tool for approximation lower bounds is a \defn{gap reduction},
which maps yes- and no-instances of a decision problem to optimization
instances with separated ranges of optimum values.
If a polynomial-time gap reduction
from 3SAT produces instances of size $N\le(n+m)^{b+o(1)}$, then
Theorem~\ref{thm:transfer} rules out $O(2^{N^{1/b-\epsilon}})$-time algorithms
that distinguish the yes- and no-cases, for every constant $\epsilon>0$, assuming SH.
For example, if a minimization problem has yes-case optimum $\leq A$
and no-case optimum $\geq B$, where $0<A<B$, then there is no
$\bigl(<\frac{B}{A}\bigr)$-approximation in the stated time.

A source of such gaps is Dinur's nearly linear PCP theorem \cite{Din07}.
As formulated by Manurangsi \cite[Theorem~6]{Man26}, it maps a 3CNF
formula with $n$ variables and $m$ clauses to another 3CNF formula with
$m'=O((n+m)\log^c(n+m))$ clauses, for a constant $c>0$,
preserving satisfiability while ensuring that every assignment violates
at least a fixed fraction $\rho>0$ of clauses in the no-case.
Applying Theorem~\ref{thm:transfer} with $b=1$ to Dinur's PCP theorem gives
the following corollary, previously observed under Weak ETH by
Manurangsi \cite[Section~2.2]{Man26}:

\begin{corollary} [Max-3SAT Gap Hardness]\label{cor:gap}
  Assume SH.
  There is a constant $0<\rho<1$ such that, for every constant $\epsilon>0$,
  no $O(2^{m^{1-\epsilon}})$-time algorithm distinguishes
  satisfiable 3CNF formulas from 3CNF formulas
  for which at most $(1-\rho)m$ out of the $m$ clauses can be satisfied.
  Consequently, for every constant $\epsilon>0$ and
  every fixed $1\le r<1/(1-\rho)$, Max-3SAT has no $r$-approximation
  in $O(2^{m^{1-\epsilon}})$ time.
\end{corollary}

The polylogarithmic blowup in Dinur's PCP theorem is absorbed by the
$\epsilon$ loss in the exponent.
For subexponential gap reductions, a polynomial-time algorithm
distinguishing the yes- and no-cases likewise contradicts Lemma~\ref{lem:size} whenever
the reduction takes $2^{O((n+m)^{\alpha})}$ time for fixed $0<\alpha<1$.

Manurangsi's subexponential reductions for Densest $k$-Subgraph
illustrate the framework.
Their earlier result \cite{Man17} ruled out polynomial-time
$n^{1/(\log\log n)^c}$-approximation on
$n$-vertex graphs under ETH, for some universal constant $c>0$.
By Theorem~\ref{thm:sh-weak-eth}, Manurangsi's \cite[Theorem~1]{Man26}
result under Weak ETH also holds under SH: it rules out
polynomial-time $2^{(\log n)^{1-\epsilon}}$-approximation for Densest
$k$-Subgraph on $n$-vertex graphs, for every constant $\epsilon>0$,
even when the yes-instance contains a $k$-clique.

\section{Counting Under \#SH \label{sec:counting}}

Counting lower bounds follow the same strategy, using counting reductions
and the size formulation of \#SH already proved in Lemma~\ref{lem:size}.
The algorithms refuting ETH \cite{OpenAI26,Anthropic26} also refute \#ETH.

One thing that sets counting apart from decision is the existence of
``easy-to-decide, hard-to-count'' problems such as \#2SAT.
Hoffmann \cite[Corollary~1.3]{Hof10} proved a
$2^{\Omega(n/\log^3 n)}$ lower bound for counting independent sets,
and hence \#2SAT, under \#ETH.
Dell et al.\ \cite[Theorem~1.2]{DHMTW14} proved tight bounds under randomized ETH.
Curticapean's thesis \cite[Corollary~8.17]{Cur15} obtained tight bounds for
sparse \#2SAT using block interpolation, ruling out $2^{o(n)}$ time under
\#ETH even for monotone 2CNF formulas in which each variable occurs
only $O(1)$ times. 

Bannach, Demaine, Gomez, and Hecher \cite{BDGH25} give a linear-size reduction
from \#SAT to the difference of two \#2SAT counts, simplified in \cite{BDGH26}.
The size formulation of
\#SH lets us apply these reductions directly.

\begin{corollary} [\#2SAT Lower Bound]\label{cor:counting-2sat}
  Assuming \#SH, for every constant $\epsilon>0$, \#2SAT on
  $n$ variables and $m$ clauses cannot be solved in
  $O(2^{(n+m)^{1-\epsilon}})$ time,
  even when restricted to formulas of constant degree.
\end{corollary}

\begin{proof}
  By decreasing the power saving if necessary, assume $0<\epsilon<1$.
  Given a 3CNF formula $F$ with $n$ variables and $m$ clauses, the
  construction of \cite[Section~3, Theorem~3.8]{BDGH26} produces two
  2CNF formulas $G_1,G_2$ in polynomial time such that
  $$
    \#F=\#G_1-\#G_2, \qquad n_i+m_i=O(n+m)\quad \text{for } i\in \{1,2\},
  $$
  where $n_i,m_i$ count the variables and clauses of $G_i$.
  An algorithm with the claimed running time would evaluate both counts
  in $O(2^{(n+m)^{1-\epsilon/2}})$ time, absorbing the constant-factor
  blowup into the power saving. Subtracting them takes polynomial
  time because they have $O(n+m)$ bits. This contradicts
  Lemma~\ref{lem:size}. The extension to constant degree is given by~\cite[Lemma~4]{BDGH25}.
\end{proof}

This argument uses two counting queries and subtraction, extending the
many-one decision reductions of Theorem~\ref{thm:transfer}. Alternatively, a single \#2SAT query on a monotone formula, followed by arithmetic postprocessing, suffices~\cite[Theorem~12]{BDGH25}.

\section{Randomization \label{sec:randomization}}

Our equivalences and reduction framework also apply to bounded-error (Monte Carlo)
randomized algorithms under the corresponding randomized hypotheses,
analogous to rETH \cite{DHMTW14}:

\begin{hypothesis}
  \defn{Randomized Satisfiability Hypothesis (rSH)}.
  For every constant $\epsilon>0$, no randomized algorithm decides
  CNF SAT on inputs of length $L$ correctly with probability at least $\frac{2}{3}$
  (over its coin flips) in $O(2^{L^{1-\epsilon}})$ time.
\end{hypothesis}

\begin{hypothesis}
  \defn{Randomized Counting Satisfiability Hypothesis (r\#SH)}.
  For every constant $\epsilon>0$, no randomized algorithm counts
  the satisfying assignments of a CNF formula of length $L$ correctly
  with probability at least $\frac{2}{3}$ (over its coin flips) in
  $O(2^{L^{1-\epsilon}})$ time.
\end{hypothesis}

Similarly define \defn{Weak rETH}, \defn{Weak $k$-rETH}, and \defn{Weak r\#ETH}.

\begin{theorem} [Randomized Extensions]\label{thm:randomized}
  \renewcommand{\labelenumi}{(\alph{enumi})}
  For bounded-error randomized algorithms:

  \begin{enumerate}
  \item For every fixed $k\ge3$, rSH, Weak rETH, and Weak $k$-rETH are equivalent;
     and r\#SH and Weak r\#ETH are equivalent
     (randomized analogs of Theorem~\ref{thm:sh-weak-eth} and
     Corollary~\ref{cor:fixed-width}).
  \item Assuming rSH, the reduction framework rules out
     $2^{O(N^{1/b-\epsilon})}L^{O(1)}$-time algorithms for every constant
     $\epsilon>0$ under the same deterministic reduction hypotheses
     (randomized analog of Theorem~\ref{thm:transfer}).
     The gap-hardness and approximation bounds for Max-3SAT also hold
     (randomized analog of Corollary~\ref{cor:gap}).
  \item Assuming r\#SH, for every constant $\epsilon>0$, \#2SAT on $n$ variables
     and $m$ clauses cannot be solved in $O(2^{(n+m)^{1-\epsilon}})$ time,
     even for formulas of constant degree
     (randomized analog of Corollary~\ref{cor:counting-2sat}).
  \end{enumerate}
\end{theorem}

\begin{proof}
  \textbf{(a)}
  The deterministic reduction in Lemma~\ref{lem:size} preserves
  the success probability of a single randomized call.
  For Corollary~\ref{cor:size}, sparsification produces
  $t=2^{O(n^{1-\delta})}$ formulas.
  Amplify each call to error at most $\frac{1}{3t}$ by $O(\log(t+1))$
  independent repetitions, taking the majority answer.
  A union bound gives probability at least $\frac{2}{3}$ that all $t$ answers
  are correct, so their disjunction is correct as well.
  Because $\log(t+1)=O(n^{1-\delta})$, amplification adds only a polynomial
  factor to the running time.
  The proofs of Theorem~\ref{thm:sh-weak-eth} and Corollary~\ref{cor:fixed-width}
  therefore give the claimed randomized decision equivalences.
  For counting, the reduction in Lemma~\ref{lem:size} is parsimonious.
  The same amplification works by taking the most frequent output value.
  With probability at least $\frac{2}{3}$, all counts are correct; disjointness
  lets us sum them to obtain the original count.
  The same argument therefore gives the equivalence of r\#SH and Weak r\#ETH.

  \textbf{(b)}
  Theorem~\ref{thm:transfer} and Corollary~\ref{cor:gap} use deterministic
  reductions followed by one call, preserving its success probability.

  \textbf{(c)}
  For Corollary~\ref{cor:counting-2sat}, amplify each of the two counts to error
  at most $\frac{1}{6}$ using $O(1)$ repetitions.
  Both counts, and hence their difference, are correct with probability
  at least $\frac{2}{3}$.
  The deterministic reduction to constant-degree formulas preserves this
  conclusion. These amplifications preserve all stated running-time bounds.
\end{proof}

\section{Line of Succession \label{sec:succession}}

Recent conjecture-breaking algorithms encourage us to plan fallbacks
in case SH fails.
Suppose CNF SAT on inputs of length $L$ admits a $2^{O(L^{\alpha})}$-time
algorithm for some fixed $\alpha$ with $0<\alpha<1$.
We could still conjecture that its power index is at least a
constant $\beta$ with $0<\beta\le\alpha$.
The weaker assertion of \emph{positive power index}
matches Lokshtanov's Weak ETH \cite{Lok17}
and rules out quasipolynomial-time algorithms.

If CNF SAT ends up having power index $0$, we can switch to finer measures.
Define the \defn{polylogarithmic power index} of a computational problem
to be the infimum over real $c\ge1$ for which the problem admits a
$2^{O((\log L)^c)}$-time algorithm.
If there is no quasipolynomial-time algorithm, this infimum is~$\infty$.
A finite polylogarithmic power index for CNF SAT would imply
$\mathrm{NP}\subseteq\mathrm{QP}$, because polynomial blowup changes
$(\log L)^c$ by only a constant factor.
$\mathrm{P}=\mathrm{NP}$ holds exactly when CNF SAT has
polylogarithmic power index $1$ and the infimum is attained,
giving a $2^{O(\log L)}=L^{O(1)}$-time algorithm.
In that case, we could study the exponent in the polynomial running time.


\section*{Acknowledgments}

We thank Ryan Williams for helpful discussions.
This paper was typeset with Texlish [\protect\url{https://texlish.org/}].

The main ideas of this paper originated from a supercollaborative
discussion among the authors.  We used OpenAI's GPT 6.1 Sol to assist
with drafting and revising the exposition, finding related literature,
and developing and checking proof details.
We reviewed and thoroughly revised the manuscript and
hold ourselves responsible for its mathematical content and writing quality.

\AddToHook
   {cmd/thebibliography/after}%
   {\setlength\itemsep{0pt}%
    \setlength\parskip{0pt}}%
\bibliographystyle{alpha}%
\bibliography{paper}%

@incollection{Tse70,
  author = {G. S. Tseitin},
  title = {On the Complexity of Derivation in Propositional Calculus},
  booktitle = {Studies in Constructive Mathematics and Mathematical Logic, Part II},
  editor = {A. O. Slisenko},
  pages = {115--125},
  publisher = {Consultants Bureau},
  year = {1970},
  doi = {10.1007/978-1-4899-5327-8_25},
  note = {Originally published in Russian in 1968.
    \url{https://doi.org/10.1007/978-1-4899-5327-8_25}}
}

@inproceedings{Coo71,
  author = {Stephen A. Cook},
  title = {The Complexity of Theorem-Proving Procedures},
  booktitle = {Proceedings of the Third Annual ACM Symposium on Theory of Computing},
  pages = {151--158},
  publisher = {ACM},
  year = {1971},
  doi = {10.1145/800157.805047},
  note = {\url{https://doi.org/10.1145/800157.805047}}
}

@incollection{Kar72,
  author = {Richard M. Karp},
  title = {Reducibility among Combinatorial Problems},
  booktitle = {Complexity of Computer Computations},
  pages = {85--103},
  publisher = {Plenum Press},
  year = {1972},
  doi = {10.1007/978-1-4684-2001-2_9},
  note = {\url{https://doi.org/10.1007/978-1-4684-2001-2_9}}
}

@article{Lev73,
  author = {Leonid A. Levin},
  title = {Universal Sequential Search Problems},
  journal = {Problems of Information Transmission},
  volume = {9},
  number = {3},
  pages = {265--266},
  year = {1973},
  note = {Translated from Problemy Peredachi Informatsii 9(3):115--116.
    \url{https://www.mathnet.ru/eng/ppi914}}
}

@article{Ward26,
  author = {Isabella Ward},
  title = {An {AI} Breakthrough Has Made Decades of Computer Science Research `Irrelevant'},
  journal = {WIRED},
  year = {2026},
  month = oct,
  note = {October 29, 2026.
    \url{https://www.wired.com/story/ai-made-decades-of-computer-science-research-irrelevant/}}
}

@article{SH90,
  author = {Richard E. Stearns and Hunt, III, Harry B.},
  title = {Power indices and easier hard problems},
  journal = {Mathematical Systems Theory},
  volume = {23},
  pages = {209--225},
  year = {1990},
  month = dec,
  doi = {10.1007/BF02090776},
  note = {\url{https://doi.org/10.1007/BF02090776}}
}

@inproceedings{IP99,
  author = {Russell Impagliazzo and Ramamohan Paturi},
  title = {Complexity of {$k$-SAT}},
  booktitle = {Proceedings of the Fourteenth Annual IEEE Conference
    on Computational Complexity},
  pages = {237--240},
  publisher = {IEEE Computer Society},
  year = {1999},
  doi = {10.1109/CCC.1999.766282},
  note = {\url{https://doi.org/10.1109/CCC.1999.766282}}
}

@article{IP01,
  author = {Russell Impagliazzo and Ramamohan Paturi},
  title = {On the Complexity of {$k$-SAT}},
  journal = {Journal of Computer and System Sciences},
  volume = {62},
  number = {2},
  pages = {367--375},
  year = {2001},
  doi = {10.1006/jcss.2000.1727},
  note = {\url{https://doi.org/10.1006/jcss.2000.1727}}
}

@article{IPZ01,
  author = {Russell Impagliazzo and Ramamohan Paturi and Francis Zane},
  title = {Which Problems Have Strongly Exponential Complexity?},
  journal = {Journal of Computer and System Sciences},
  volume = {63},
  number = {4},
  pages = {512--530},
  year = {2001},
  doi = {10.1006/jcss.2001.1774},
  note = {\url{https://doi.org/10.1006/jcss.2001.1774}}
}

@inproceedings{CIP06,
  author = {Chris Calabro and Russell Impagliazzo and Ramamohan Paturi},
  title = {A Duality between Clause Width and Clause Density for {SAT}},
  booktitle = {Proceedings of the 21st Annual IEEE Conference on Computational Complexity},
  pages = {252--260},
  publisher = {IEEE Computer Society},
  year = {2006},
  doi = {10.1109/CCC.2006.6},
  note = {\url{https://doi.org/10.1109/CCC.2006.6}}
}

@misc{OpenAI26,
  author = {{OpenAI}},
  title = {The {Exponential Time Hypothesis} is False},
  year = {2026},
  month = oct,
  note = {Dated October 17, revised October 25, 2026.
    \url{https://github.com/openai/math/tree/main/preprints/The-Exponential-Time-Hypothesis-is-False-October-25-2026}.
    Publicly announced October 26, 2026:
    \url{https://openai.com/index/building-scientific-understanding-with-academic-communities/}}
}

@misc{Anthropic26,
  author = {{Anthropic}},
  title = {A {$2^{o(n)}$} Time Algorithm for {3-SAT}},
  year = {2026},
  month = oct,
  howpublished = {Hexagon:2610.00577v1},
  note = {Dated October 26, 2026.
    \url{https://hexagonmath.org/2610.00577v1}}
}

@misc{Lok17,
  author = {Daniel Lokshtanov},
  title = {Answer to: Limits of variants of {Independent Set}?},
  howpublished = {Theoretical Computer Science Stack Exchange},
  year = {2017},
  month = may,
  note = {Posted as \texttt{daniello}, May 25, 2017.
    \url{https://cstheory.stackexchange.com/questions/38283/limits-of-variants-of-independent-set}}
}

@inproceedings{CHIKK21,
  author = {Marco Carmosino and Kenneth Hoover and Russell Impagliazzo
    and Valentine Kabanets and Antonina Kolokolova},
  title = {Lifting for Constant-Depth Circuits and Applications to {MCSP}},
  booktitle = {48th International Colloquium on Automata, Languages, and Programming},
  series = {Leibniz International Proceedings in Informatics},
  volume = {198},
  pages = {44:1--44:20},
  publisher = {Schloss Dagstuhl--Leibniz-Zentrum f{\"u}r Informatik},
  year = {2021},
  doi = {10.4230/LIPIcs.ICALP.2021.44},
  note = {\url{https://doi.org/10.4230/LIPIcs.ICALP.2021.44}}
}

@article{LMS18,
  author = {Daniel Lokshtanov and D{\'a}niel Marx and Saket Saurabh},
  title = {Slightly Superexponential Parameterized Problems},
  journal = {SIAM Journal on Computing},
  volume = {47},
  number = {3},
  pages = {675--702},
  year = {2018},
  doi = {10.1137/16M1104834},
  note = {\url{https://doi.org/10.1137/16M1104834}}
}

@misc{Man26,
  author = {Pasin Manurangsi},
  title = {Almost-Polynomial Ratio {Almost-ETH}-Hardness of Approximating {Densest $k$-Subgraph}},
  year = {2026},
  month = oct,
  howpublished = {arXiv:1611.05991v3},
  note = {Version 3, posted October 27, 2026.
    \url{https://arxiv.org/abs/1611.05991v3}}
}

@article{Din07,
  author = {Irit Dinur},
  title = {The {PCP} Theorem by Gap Amplification},
  journal = {Journal of the ACM},
  volume = {54},
  number = {3},
  articleno = {12},
  year = {2007},
  doi = {10.1145/1236457.1236459},
  note = {\url{https://www.wisdom.weizmann.ac.il/~dinuri/mypapers/combpcp.pdf}}
}

@article{LMS11,
  author = {Daniel Lokshtanov and D{\'a}niel Marx and Saket Saurabh},
  title = {Lower Bounds Based on the {Exponential Time Hypothesis}},
  journal = {Bulletin of the EATCS},
  volume = {105},
  pages = {41--72},
  year = {2011},
  note = {\url{https://sites.cs.ucsb.edu/~daniello/papers/surveyETH.pdf}}
}

@article{CJ03,
  author = {Liming Cai and David W. Juedes},
  title = {On the existence of subexponential parameterized algorithms},
  journal = {Journal of Computer and System Sciences},
  volume = {67},
  number = {4},
  pages = {789--807},
  year = {2003},
  doi = {10.1016/S0022-0000(03)00074-6},
  note = {\url{https://doi.org/10.1016/S0022-0000(03)00074-6}}
}

@article{Lic82,
  author = {David Lichtenstein},
  title = {Planar Formulae and Their Uses},
  journal = {SIAM Journal on Computing},
  volume = {11},
  number = {2},
  pages = {329--343},
  year = {1982},
  doi = {10.1137/0211025},
  note = {\url{https://doi.org/10.1137/0211025}}
}

@article{BG19,
  author = {{\'E}douard Bonnet and Panos Giannopoulos},
  title = {Orthogonal Terrain Guarding is {NP}-complete},
  journal = {Journal of Computational Geometry},
  volume = {10},
  number = {2},
  pages = {21--44},
  year = {2019},
  doi = {10.20382/jocg.v10i2a3},
  note = {\url{https://doi.org/10.20382/jocg.v10i2a3}}
}

@article{BBKMSZ20,
  author = {Mark de Berg and Hans L. Bodlaender and S{\'a}ndor Kisfaludi-Bak
    and D{\'a}niel Marx and Tom C. van der Zanden},
  title = {A Framework for {Exponential-Time-Hypothesis}--Tight Algorithms
    and Lower Bounds in Geometric Intersection Graphs},
  journal = {SIAM Journal on Computing},
  volume = {49},
  number = {6},
  pages = {1291--1331},
  year = {2020},
  doi = {10.1137/20M1320870},
  note = {\url{https://doi.org/10.1137/20M1320870}}
}

@misc{Man17,
  author = {Pasin Manurangsi},
  title = {Almost-Polynomial Ratio {ETH}-Hardness of Approximating {Densest $k$-Subgraph}},
  year = {2017},
  month = apr,
  howpublished = {arXiv:1611.05991v2},
  note = {Version 2, posted April 10, 2017.
    \url{https://arxiv.org/abs/1611.05991v2}}
}

@article{DHMTW14,
  author = {Holger Dell and Thore Husfeldt and D{\'a}niel Marx
    and Nina Taslaman and Martin Wahl{\'e}n},
  title = {Exponential Time Complexity of the Permanent and the {Tutte} Polynomial},
  journal = {ACM Transactions on Algorithms},
  volume = {10},
  pages = {21:1--21:32},
  year = {2014},
  doi = {10.1145/2635812},
  note = {\url{https://arxiv.org/abs/1206.1775}}
}

@inproceedings{Hof10,
  author = {Christian Hoffmann},
  title = {Exponential Time Complexity of Weighted Counting of Independent Sets},
  booktitle = {Parameterized and Exact Computation:
    5th International Symposium (IPEC 2010)},
  series = {Lecture Notes in Computer Science},
  volume = {6478},
  pages = {180--191},
  publisher = {Springer},
  year = {2010},
  doi = {10.1007/978-3-642-17493-3_18},
  note = {\url{https://arxiv.org/abs/1007.1146}}
}

@inproceedings{BDGH25,
  author = {Max Bannach and Erik D. Demaine and Timothy Gomez and Markus Hecher},
  title = {{{\#P}} is Sandwiched by One and Two {{\#2DNF}} Calls:
    Is Subtraction Stronger Than We Thought?},
  booktitle = {40th Annual ACM/IEEE Symposium on Logic in Computer Science (LICS)},
  pages = {31--43},
  publisher = {IEEE},
  year = {2025},
  doi = {10.1109/LICS65433.2025.00010},
  note = {\url{https://doi.org/10.1109/LICS65433.2025.00010}}
}

@inproceedings{BDGH26,
  author = {Max Bannach and Erik D. Demaine and Timothy Gomez and Markus Hecher},
  title = {A Novel Reduction from {{\#SAT}} to {{\#2SAT}} Based on Symmetry:
    Simply Drop the Large Clauses},
  booktitle = {2026 SIAM Symposium on Simplicity in Algorithms (SOSA)},
  pages = {254--265},
  publisher = {SIAM},
  year = {2026},
  doi = {10.1137/1.9781611978964.19},
  note = {\url{https://doi.org/10.1137/1.9781611978964.19}}
}

@misc{Sze26,
  author = {Stefan Szeider},
  title = {After {ETH}: the {$k\times k$ Clique Hypothesis}},
  year = {2026},
  month = oct,
  howpublished = {Zenodo technical note, version 1},
  doi = {10.5281/zenodo.23797197},
  note = {Dated October 30, 2026.
    \url{https://doi.org/10.5281/zenodo.23797197}}
}

@misc{VWZ26,
  author = {Virginia {Vassilevska Williams} and Or Zamir},
  title = {The {$k$-Clique} Hypothesis Is Dead, for Good},
  year = {2026},
  month = nov,
  howpublished = {arXiv:2611.09764},
  note = {\url{https://arxiv.org/abs/2611.09764}}
}

@misc{UCSD26,
  author = {{ETH Group, UC San Diego}},
  title = {Weak {ETH} and fixed clause width: {A} proof by quantitative sparsification},
  year = {2026},
  month = oct,
  howpublished = {ETH Repository},
  note = {Dated October 28, 2026.
    \url{https://drive.google.com/file/d/1Pfj100VaWs63cwC-AqzUaOgz1S0lGU59/view}}
}

@misc{Aar26,
  author = {Scott Aaronson},
  title = {Comment {\#31} on {Podcasts about the Topic: Part II}},
  year = {2026},
  month = oct,
  howpublished = {Shtetl-Optimized},
  note = {October 27, 2026, 6:29 a.m.
    \url{https://scottaaronson.blog/?p=10239#comment-2042469}}
}

@misc{Lam26a,
  author = {Michael Lampis},
  title = {Answer to: {How} to salvage arguments that assume {ETH}},
  year = {2026},
  month = oct,
  howpublished = {Theoretical Computer Science Stack Exchange},
  note = {Answer 57313, October 27, 2026.
    \url{https://cstheory.stackexchange.com/a/57313}}
}

@misc{Lam26b,
  author = {Michael Lampis},
  title = {Comment {\#55} on {Podcasts about the Topic: Part II}},
  year = {2026},
  month = oct,
  howpublished = {Shtetl-Optimized},
  note = {October 27, 2026, 4:15 p.m.
    \url{https://scottaaronson.blog/?p=10239#comment-2042514}}
}

@misc{Li26,
  author = {Xiaoyu Li},
  title = {What Does the Refutation of the {Exponential Time Hypothesis}
    Mean for Learning Theory?},
  year = {2026},
  month = oct,
  howpublished = {Hexagon:2610.00613v1},
  note = {Posted October 29, 2026.
    \url{https://hexagonmath.org/2610.00613v1}}
}

@article{CRTY23,
  author = {Lijie Chen and Ron D. Rothblum and Roei Tell and Eylon Yogev},
  title = {On Exponential-time Hypotheses, Derandomization,
    and Circuit Lower Bounds},
  journal = {Journal of the ACM},
  volume = {70},
  number = {4},
  pages = {25:1--25:62},
  year = {2023},
  doi = {10.1145/3593581},
  note = {\url{https://doi.org/10.1145/3593581}}
}

@phdthesis{Cur15,
  author = {Radu Curticapean},
  title = {The Simple, Little and Slow Things Count:
    On Parameterized Counting Complexity},
  school = {Saarland University},
  year = {2015},
  doi = {10.22028/D291-26612},
  note = {\url{https://doi.org/10.22028/D291-26612}}
}

@inproceedings{BKN23,
  author = {Benjamin Bergougnoux and Tuukka Korhonen and Jesper Nederlof},
  title = {Tight Lower Bounds for Problems Parameterized by Rank-Width},
  booktitle = {40th International Symposium on Theoretical Aspects of
    Computer Science (STACS 2023)},
  series = {Leibniz International Proceedings in Informatics},
  volume = {254},
  pages = {11:1--11:17},
  publisher = {Schloss Dagstuhl--Leibniz-Zentrum f{\"u}r Informatik},
  year = {2023},
  doi = {10.4230/LIPIcs.STACS.2023.11},
  note = {\url{https://doi.org/10.4230/LIPIcs.STACS.2023.11}}
}

@article{Bon26,
  author = {{\'E}douard Bonnet},
  title = {Treewidth Inapproximability and Tight {ETH} Lower Bound},
  journal = {Journal of the ACM},
  volume = {73},
  number = {5},
  pages = {35:1--35:10},
  year = {2026},
  doi = {10.1145/3833387},
  note = {\url{https://doi.org/10.1145/3833387}}
}

@article{CFGK17,
  author = {Marek Cygan and Fedor V. Fomin and Alexander Golovnev
    and Alexander S. Kulikov and Ivan Mihajlin and Jakub Pachocki
    and Arkadiusz Soca{\l}a},
  title = {Tight Lower Bounds on Graph Embedding Problems},
  journal = {Journal of the ACM},
  volume = {64},
  number = {3},
  pages = {18:1--18:22},
  year = {2017},
  doi = {10.1145/3051094},
  note = {\url{https://doi.org/10.1145/3051094}}
}

@inproceedings{BZ17,
  author = {Hans L. Bodlaender and Tom C. van der Zanden},
  title = {Improved Lower Bounds for Graph Embedding Problems},
  booktitle = {Algorithms and Complexity: 10th International Conference
    (CIAC 2017)},
  series = {Lecture Notes in Computer Science},
  volume = {10236},
  pages = {92--103},
  publisher = {Springer},
  year = {2017},
  doi = {10.1007/978-3-319-57586-5_9},
  note = {\url{https://doi.org/10.1007/978-3-319-57586-5_9}}
}

@inproceedings{Sch78,
  author = {Thomas J. Schaefer},
  title = {The Complexity of Satisfiability Problems},
  booktitle = {Proceedings of the Tenth Annual ACM Symposium on Theory of Computing},
  pages = {216--226},
  publisher = {ACM},
  year = {1978},
  doi = {10.1145/800133.804350},
  note = {\url{https://doi.org/10.1145/800133.804350}}
}

@article{MR08,
  author = {Wolfgang Mulzer and G{\"u}nter Rote},
  title = {Minimum-Weight Triangulation is {NP}-Hard},
  journal = {Journal of the ACM},
  volume = {55},
  number = {2},
  pages = {11:1--11:29},
  year = {2008},
  doi = {10.1145/1346330.1346336},
  note = {\url{https://doi.org/10.1145/1346330.1346336}}
}

@book{DGH26,
  author = {Erik D. Demaine and William Gasarch and Mohammad T. Hajiaghayi},
  title = {Computational Intractability: A Guide to Algorithmic Lower Bounds},
  publisher = {MIT Press},
  year = {2026},
  note = {\url{https://hardness.mit.edu/}}
}

@article{Jea98,
  author = {Peter Jeavons},
  title = {On the Algebraic Structure of Combinatorial Problems},
  journal = {Theoretical Computer Science},
  volume = {200},
  number = {1--2},
  pages = {185--204},
  year = {1998},
  doi = {10.1016/S0304-3975(97)00230-2},
  note = {\url{https://doi.org/10.1016/S0304-3975(97)00230-2}}
}

@article{CDL+16,
  author = {Marek Cygan and Holger Dell and Daniel Lokshtanov and D{\'a}niel Marx
    and Jesper Nederlof and Yoshio Okamoto and Ramamohan Paturi
    and Saket Saurabh and Magnus Wahlstr{\"o}m},
  title = {On Problems as Hard as {CNF-SAT}},
  journal = {ACM Transactions on Algorithms},
  volume = {12},
  number = {3},
  pages = {41:1--41:24},
  year = {2016},
  doi = {10.1145/2925416},
  note = {Extended abstract in CCC 2012.
    \url{https://arxiv.org/abs/1112.2275}}
}

\end{document}